\RequirePackage{lineno}
\documentclass[twocolumn,aps,prc,showpacs,superscriptaddress,floatfix]{revtex4}
\usepackage{epsfig}
\setpagewiselinenumbers
\linenumbers

\newcommand {\snn}	{\sqrt{s_{_{\rm NN}}}}
\newcommand {\gev}	{{GeV/$c$}}
\newcommand {\pp}	{{$p$+$p$}}
\newcommand {\dAu}	{{$d$+Au}}
\newcommand {\zyam}	{ZYAM}
\newcommand {\etal}	{{\it et al.}}
\newcommand {\pt}	{p_{T}}
\newcommand {\ptt}	{p_{T}^{(t)}}
\newcommand {\pta}	{p_{T}^{(a)}}
\newcommand {\phis}	{\phi_{s}}
\newcommand {\phit}	{\phi_{t}}
\newcommand {\psiEP}	{\psi_{\rm EP}}
\newcommand {\dphi}	{\Delta\phi}
\newcommand {\deta}	{\Delta\eta}
\newcommand {\vt}	{v^{(t)}}
\newcommand {\va}	{v^{(a)}}
\newcommand {\vtR}	{v^{(t,\phis)}}
\newcommand {\ff}[2]	{v_{#1}\{#2\}}
\newcommand {\fff}[3]	{v_{#1}^{(#2)}\{#3\}}
\newcommand {\VVuc}     {V_4\{{\rm uc}\}}
\newcommand {\etagap}	{\eta_{\rm gap}}
\newcommand {\mean}[1]	{\langle #1\rangle}

\begin{document}
\title{Event-plane dependent dihadron correlations with harmonic $v_n$ subtraction in Au+Au Collisions at $\snn=200$~GeV}
\affiliation{Argonne National Laboratory, Argonne, Illinois 60439, USA}
\affiliation{Brookhaven National Laboratory, Upton, New York 11973, USA}
\affiliation{University of California, Berkeley, California 94720, USA}
\affiliation{University of California, Davis, California 95616, USA}
\affiliation{University of California, Los Angeles, California 90095, USA}
\affiliation{Universidade Estadual de Campinas, Sao Paulo, Brazil}
\affiliation{University of Illinois at Chicago, Chicago, Illinois 60607, USA}
\affiliation{Creighton University, Omaha, Nebraska 68178, USA}
\affiliation{Czech Technical University in Prague, FNSPE, Prague, 115 19, Czech Republic}
\affiliation{Nuclear Physics Institute AS CR, 250 68 \v{R}e\v{z}/Prague, Czech Republic}
\affiliation{University of Frankfurt, Frankfurt, Germany}
\affiliation{Institute of Physics, Bhubaneswar 751005, India}
\affiliation{Indian Institute of Technology, Mumbai, India}
\affiliation{Indiana University, Bloomington, Indiana 47408, USA}
\affiliation{Alikhanov Institute for Theoretical and Experimental Physics, Moscow, Russia}
\affiliation{University of Jammu, Jammu 180001, India}
\affiliation{Joint Institute for Nuclear Research, Dubna, 141 980, Russia}
\affiliation{Kent State University, Kent, Ohio 44242, USA}
\affiliation{University of Kentucky, Lexington, Kentucky, 40506-0055, USA}
\affiliation{Institute of Modern Physics, Lanzhou, China}
\affiliation{Lawrence Berkeley National Laboratory, Berkeley, California 94720, USA}
\affiliation{Massachusetts Institute of Technology, Cambridge, MA 02139-4307, USA}
\affiliation{Max-Planck-Institut f\"ur Physik, Munich, Germany}
\affiliation{Michigan State University, East Lansing, Michigan 48824, USA}
\affiliation{Moscow Engineering Physics Institute, Moscow Russia}
\affiliation{NIKHEF and Utrecht University, Amsterdam, The Netherlands}
\affiliation{Ohio State University, Columbus, Ohio 43210, USA}
\affiliation{Old Dominion University, Norfolk, VA, 23529, USA}
\affiliation{Panjab University, Chandigarh 160014, India}
\affiliation{Pennsylvania State University, University Park, Pennsylvania 16802, USA}
\affiliation{Institute of High Energy Physics, Protvino, Russia}
\affiliation{Purdue University, West Lafayette, Indiana 47907, USA}
\affiliation{Pusan National University, Pusan, Republic of Korea}
\affiliation{University of Rajasthan, Jaipur 302004, India}
\affiliation{Rice University, Houston, Texas 77251, USA}
\affiliation{Universidade de Sao Paulo, Sao Paulo, Brazil}
\affiliation{University of Science \& Technology of China, Hefei 230026, China}
\affiliation{Shandong University, Jinan, Shandong 250100, China}
\affiliation{Shanghai Institute of Applied Physics, Shanghai 201800, China}
\affiliation{SUBATECH, Nantes, France}
\affiliation{Texas A\&M University, College Station, Texas 77843, USA}
\affiliation{University of Texas, Austin, Texas 78712, USA}
\affiliation{Tsinghua University, Beijing 100084, China}
\affiliation{United States Naval Academy, Annapolis, MD 21402, USA}
\affiliation{Valparaiso University, Valparaiso, Indiana 46383, USA}
\affiliation{Variable Energy Cyclotron Centre, Kolkata 700064, India}
\affiliation{Warsaw University of Technology, Warsaw, Poland}
\affiliation{University of Washington, Seattle, Washington 98195, USA}
\affiliation{Wayne State University, Detroit, Michigan 48201, USA}
\affiliation{Institute of Particle Physics, CCNU (HZNU), Wuhan 430079, China}
\affiliation{Yale University, New Haven, Connecticut 06520, USA}
\affiliation{University of Zagreb, Zagreb, HR-10002, Croatia}

\author{H.~Agakishiev}\affiliation{Joint Institute for Nuclear Research, Dubna, 141 980, Russia}
\author{M.~M.~Aggarwal}\affiliation{Panjab University, Chandigarh 160014, India}
\author{Z.~Ahammed}\affiliation{Lawrence Berkeley National Laboratory, Berkeley, California 94720, USA}
\author{A.~V.~Alakhverdyants}\affiliation{Joint Institute for Nuclear Research, Dubna, 141 980, Russia}
\author{I.~Alekseev~~}\affiliation{Alikhanov Institute for Theoretical and Experimental Physics, Moscow, Russia}
\author{J.~Alford}\affiliation{Kent State University, Kent, Ohio 44242, USA}
\author{B.~D.~Anderson}\affiliation{Kent State University, Kent, Ohio 44242, USA}
\author{C.~D.~Anson}\affiliation{Ohio State University, Columbus, Ohio 43210, USA}
\author{D.~Arkhipkin}\affiliation{Brookhaven National Laboratory, Upton, New York 11973, USA}
\author{G.~S.~Averichev}\affiliation{Joint Institute for Nuclear Research, Dubna, 141 980, Russia}
% ...

\collaboration{STAR Collaboration}\noaffiliation

\begin{abstract}
STAR measurements of dihadron azimuthal correlations ($\dphi$) are reported in mid-central (20-60\%) Au+Au collisions at $\snn=200$~GeV as a function of the trigger particle's azimuthal angle relative to the event plane, $\phis=|\phit-\psiEP|$. The elliptic ($v_2$), triangular ($v_3$), and quadratic ($v_4$) flow harmonic backgrounds are subtracted using the Zero Yield At Minimum (\zyam) method. The results are compared to minimum-bias d+Au collisions. It is found that a finite near-side ($|\dphi|<\pi/2$) long-range pseudorapidity correlation (ridge) is present in the in-plane direction ($\phis\sim0$). The away-side ($|\dphi|>\pi/2$) correlation shows a modification from d+Au data, varying with $\phis$. The modification may be a consequence of pathlength-dependent jet-quenching and may 
lead to a better understanding of high-density QCD.
\end{abstract}
\maketitle

The hot and dense QCD matter created in heavy-ion collisions at the Relativistic Heavy-Ion Collider (RHIC) of Brookhaven National Laboratory reveals properties of a nearly perfect fluid of strongly interacting quarks and gluons~\cite{wp}. 
These properties include strong elliptical azimuthal emission as large as hydrodynamical prediction relative to the initial geometry eccentricity~\cite{Kolb}, and strong attenuation of high transverse momentum ($\pt$) particles due to jet-medium interactions (jet-quenching)~\cite{XNWang,single}. 
The energy lost at high $\pt$ must be redistributed to lower $\pt$ particles~\cite{jetspec}. The distribution of those particles relative to a high-$\pt$ trigger particle can therefore provide information about the nature of the QCD interactions.

The magnitude of the effect from jet-medium interactions should depend on the pathlength the jet traverses~\cite{XNWang}. This pathlength dependence may be studied in non-central heavy-ion collisions~\cite{Kirill}, where the transverse overlap region between the two colliding nuclei is anisotropic. The short-axis direction of the overlap region may be estimated by the direction of the most probable particle emission~\cite{flowMethod}. The estimated direction together with the beam axis is called the event plane (EP), and is a proxy for the initial geometry participant plane ($\psi_2$)~\cite{Alver_fluc}. By selecting the trigger particle's azimuth relative to the event plane, $\phis=|\phit-\psiEP|$, one effectively selects different average pathlengths through the medium that the away-side jet traverses, providing differential information unavailable to inclusive jet-like dihadron correlation measurements. 

In this work, non-central 20-60\% Au+Au collisions at the nucleon-nucleon center of mass energy of $\snn=200$~GeV are analysed~\cite{thesis,long}. As a reference inclusive dihadron correlation data from minimum bias \dAu\ collisions, which include cold nuclear matter effects, are presented. (The minimum bias \dAu\ and \pp\ data are similar~\cite{single,jetspec}.) The Au+Au and d+Au data were taken by the STAR experiment at RHIC in 2004 and 2003, respectively. 
The details of the STAR experiment can be found in Ref.~\cite{STAR}. The main detector used for this analysis is the Time Projection Chamber (TPC)~\cite{TPC}, residing in a solenoidal magnet (0.5 Tesla magnetic field along the beam-axis). 
Events with a primary vertex within $\pm 30$~cm of the TPC center are used. The Au+Au centrality is defined by the measured charged particle multiplicity in the TPC within $|\eta|<0.5$~\cite{spec200}. 
Tracks are used if they are composed of at least 20 hits and 51\% of the maximum possible hits and extrapolate to within 2~cm of the primary vertex.
The same event and track cuts are applied to particle tracks used for event-plane reconstruction and for the correlation analysis.

Particles with $\pt<2$~\gev\ are used to determine the second-order harmonic event plane to ensure good event-plane resolutions. To avoid self-correlations, particles from the $\pt$ bin used in the correlation analysis (e.g., $1<\pta<1.5$~\gev) are excluded from EP reconstruction~\cite{long}. 
Nonflow correlations~\cite{nonflow}, such as dijets, can influence the EP determination. To reduce this effect, particles within $|\deta|=|\eta-\eta_{\rm trig}|<0.5$ from the trigger particle are excluded from the EP reconstruction in this analysis~\cite{long}. This is called the modified reaction-plane (MRP) method~\cite{v2MRP}. The traditional EP method, on the other hand, does not exclude those particles in the vicinity of the trigger particle in $\eta$. Remaining possible biases due to trigger-EP correlations may be estimated by comparing results relative to the EP reconstructed from these two methods. The results are found to be quantitatively similar which suggests that such biases may be small~\cite{long}.

Dihadron correlations are analyzed for pairs within pseudorapidity $|\eta|<1$. The trigger particle $\pt$ range is $3<\ptt<4$~\gev. Two associated particle $\pt$ bins, $1<\pta<1.5$~\gev\ and $1.5<\pta<2$~\gev, are analyzed and then added together in the final results. These choices of $\pt$ ranges are motivated by the expectation of significant jet contributions and the need for reasonable statistics~\cite{long}. The data are divided into six equal-size slices in $\phis$ and analyzed in azimuthal angle difference ($\dphi$) and pseudorapidity difference ($\deta$) between associated and trigger particle. The associated particle yields are corrected for single-particle track reconstruction efficiency which is obtained from embedding simulated tracks into real events~\cite{Levente}. 
The detector non-uniformity in $\dphi$ is corrected by the event-mixing technique, where the trigger particle from one event is paired with associated particles from another event with approximately matching primary vertex position and event multiplicity~\cite{jetspec,long}. The two-particle acceptance in $\deta$, approximately triangle-shaped, is not corrected for~\cite{jetspec}. 
The correlation function is normalized by the number of trigger particles in its corresponding $\phis$ bin. 

\begin{figure*}[hbt]
\includegraphics[width=1.03\textwidth]{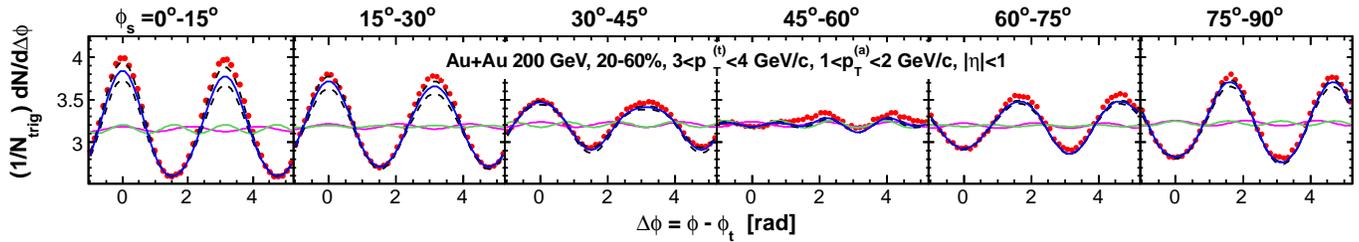}
\caption{(Color online) Raw dihadron $\dphi$ correlations (data points) as a function of $\phis=|\phit-\psiEP|$, with a cut on the trigger-associated pseudo-rapidity difference of $|\deta|>0.7$. The triangle two-particle $\deta$ acceptance is not corrected. Statistical errors are smaller than the symbol size; systematic uncertainty is 5\% (not shown). The curves are flow modulated \zyam\ background by Eq.~(\ref{eq:bkgd}) (blue solid), its systematic uncertainty boundaries (dashed), and the $v_3$ (pink solid) and $\ff{4}{\psi_2}$ (green solid) contributions.}
\label{fig:raw}
\end{figure*}

Figure~\ref{fig:raw} shows the raw azimuthal correlations as a function of $\phis$. A cut of $|\deta|>0.7$ is applied on the pseudorapidity difference between the trigger and associated particles in order to minimize the near-side jet contributions~\cite{long}. The overall systematic uncertainty on the raw correlation functions is 5\%, dominated by that in the efficiency correction.

Particles from the underlying event are uncorrelated with the trigger particle (and the corresponding jet), and follow the non-uniform distribution pattern in $\Delta\phi$ defined by the anisotropic flow. This background has to be removed in order to study jet-like correlations.
The major background contribution comes from elliptic flow ($v_2$). However, quadratic flow ($v_4$) correlated to $\psi_2$ can also have a sizable contribution~\cite{Bielcikova}. Due to fluctuations in the initial overlap geometry~\cite{Alver_fluc}, finite odd harmonic flows, particularly triangular flow ($v_3$) can also contribute~\cite{Mishra}. 
Such odd harmonics are reproduced in transport models AMPT (A Multi-Phase Transport)~\cite{Xu} and UrQMD (Ultra-relativistic Quantum Molecular Dynamics)~\cite{Petersen}, as well as in event-by-event hydrodynamic calculations with hot spots~\cite{Takahashi} or incorporating initial geometry fluctuations~\cite{Schenke}. The measured $v_3$ by both the event-plane and two-particle cumulant methods at RHIC~\cite{PHENIXv3,STARv3} are qualitatively consistent with hydrodynamic calculations. 

In this analysis, the flow correlated background is given by~\cite{Bielcikova}
\begin{widetext}
\begin{equation}
\frac{dN}{d\dphi}=B\left(1+2\va_2\vtR_2\cos2\dphi+2\va_3\vt_3\cos3\dphi+2\fff{4}{a}{\psi_2}\fff{4}{t,\phis}{\psi_2}\cos4\dphi+2\VVuc\cos4\dphi\right).\label{eq:bkgd}
\end{equation}
\end{widetext}
Here $B$ is the background normalization (see below); $\va_2$ and $\fff{4}{a}{\psi_2}$ are the associated particles' second and fourth harmonics with respect to $\psi_2$; and $\vtR_2$ and $\fff{4}{t,\phis}{\psi_2}$ are the average harmonics of the trigger particles, $\vtR_n=\mean{\cos n(\phit-\psi_2)}^{(\phis)}$, 
where the averages are taken over the slice around $\phis$ as $\phis-\pi/24<\left|\phit-\psiEP\right|<\phis+\pi/24$.
Since the triangularity orientation is random relative to $\psi_2$, the triangular flow background is independent of EP, where $\vt_3$ and $\va_3$ are the trigger and associated particle triangular flows. The last term in Eq.~(\ref{eq:bkgd}) arises from $v_4$ fluctuations uncorrelated to $\psi_2$ (see below). Higher-order harmonic flows are negligible~\cite{long}.

Eq.~(\ref{eq:bkgd}) does not include the first order harmonic, $v_1$. The effect of directed flow, rapidity-odd due to collective sidewards deflection of particles, is small and can be neglected~\cite{v1paper}. It has been suggested~\cite{Teaney} that $v_1$ fluctuation effects (sometimes called rapidity-even $v_1$) may not be small due to initial geometry fluctuations. Preliminary data~\cite{STAR_dipole} indicate that the dipole fluctuation effect changes sign at $\pt\approx1$~\gev, negative at lower $\pt$ and positive at higher $\pt$. For $\pta=1$-2~\gev\ used in this analysis, the dipole fluctuation effect is approximately zero and may be neglected. 
Note that the possible effect of statistical global momentum conservation can generate a negative dipole. However, this is considered as part of the correlation signal, just as momentum conservation by any other mechanisms, for example dijet production.

The flow correlated background given by Eq.~(\ref{eq:bkgd}) is shown in Fig.~\ref{fig:raw} as solid curves. The background curves have been normalized assuming that the background-subtracted signal has Zero Yield At Minimum (\zyam)~\cite{jetspec,zyam}. An alternative approach that has been used to describe dihadron correlation data treats the anisotropic flow modulations as free parameters in a multi-parameter model fit to the dihadron correlation functions in 2-dimensional $\deta$-$\dphi$ space~\cite{minijet2}. 
A detailed discussion can be found in Ref.~\cite{long}.

\begin{figure*}[hbt]
\includegraphics[width=1.03\textwidth]{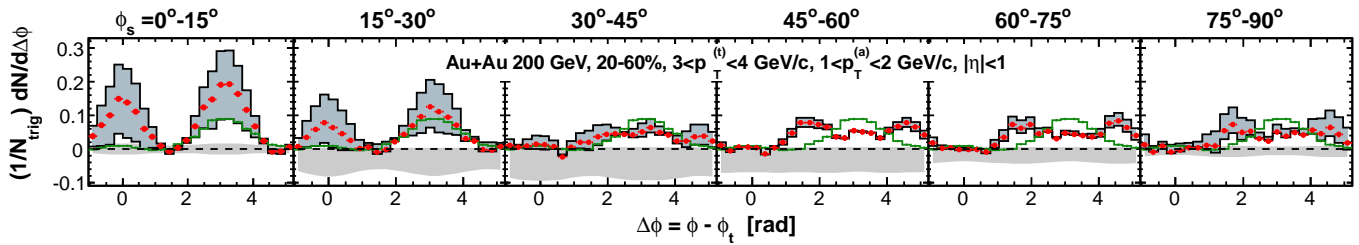}
\caption{(Color online) Background-subtracted dihadron $\dphi$ correlations as a function of $\phis=|\phit-\psiEP|$, with a cut on the trigger-associated pseudo-rapidity difference of $|\deta|>0.7$. The triangle two-particle $\deta$ acceptance is not corrected. Flow background is subtracted by Eq.~(\ref{eq:bkgd}). Systematic uncertainties due to flow subtraction are shown as black histograms enclosing the shaded area; those due to the \zyam\ normalization are shown in the horizontal shaded band around zero. Statistical errors are smaller than the point size. For comparison, the inclusive dihadron correlations from \dAu\ collisions are superimposed as the green histogram with statistical errors.}
\label{fig:corr}
\end{figure*}

The major systematic uncertainties on the results reported here come from uncertainties in the determination of the anisotropic flows. 
Two $v_2$ measurements are used~\cite{flowMethod}. One is the two-particle cumulant $\ff{2}{2}$ which overestimates elliptic flow due to nonflow contaminations. 
A major component of nonflow comes from correlated pairs at small opening angle~\cite{minijet2}. 
To suppress nonflow, a pseudo-rapidity $\eta$-gap ($\etagap$) of 0.7 is applied between the particle of interest and the reference particle used in the $\ff{2}{2}$ measurement. However, away-side two-particle correlations, such as those due to di-jets, cannot be eliminated. The other measurement is the four-particle cumulant $\ff{2}{4}$ which underestimates elliptic flow because the flow fluctuation effect in $\ff{2}{4}$ is negative~\cite{Aihong}. The range between $\ff{2}{2}$ and $\ff{2}{4}$ is therefore treated as a systematic uncertainty, as in Refs.~\cite{jetspec}, and their average is used as the best estimate for $v_2$. $v_3$ and $v_4$ are obtained by the two-particle cumulant method~\cite{STARv3,long} with $\etagap=0.7$, as for $\ff{2}{2}$. Since $\ff{3}{2}$ decreases with $\deta$~\cite{STARv3}, the $\ff{3}{2}$ represents the maximum flow for the correlation functions at $|\deta|>0.7$. The $\ff{4}{\psi_2}$ is parameterized~\cite{v2MRP} by $\ff{4}{\psi_2}=1.15v_2^2$. The $\psi_2$-uncorrelated $\VVuc$ is obtained as $\sqrt{\ff{4}{2}^2-\ff{4}{\psi_2}^2}$, and is found to be negligible for the 20-60\% centrality range used in this analysis~\cite{long}. The $v_n$ values used in the flow background subtraction are tabulated in Ref.~\cite{long}.

Another major source of systematic uncertainties comes from background normalization by \zyam. This is assessed by varying the size of the normalization range in $\dphi$ between $\pi/12$ and $\pi/4$ (default is $\pi/6$), similar to what was done in Ref.~\cite{jetspec}. The \zyam\ assumption likely gives an upper limit to the background from the underlying event. To estimate this effect, two \zyam\ background levels are obtained from correlation functions at positive $\phit-\psiEP$ and negative $\phit-\psiEP$ respectively. Those \zyam\ backgrounds are always lower than the default $B$ from \zyam\ of the combined correlation function. The difference is treated as an additional, one-sided systematic uncertainty on $B$. The different sources of systematic uncertainties on $B$ are added in quadrature. 


Figure~\ref{fig:corr} shows the background-subtracted dihadron correlations as a function of $\phis$. The black histograms enclosing the shaded area indicate the systematic uncertainties due to anisotropic flow. The horizontal shaded band around zero indicates the systematic uncertainties due to \zyam\ background normalization. 
For comparison the minimum-bias \dAu\ inclusive dihadron correlation (without differentiating with respect to an ``event plane'') is superimposed in each panel in Fig.~\ref{fig:corr}. 
For both Au+Au and \dAu, a cut of $|\deta|>0.7$ is applied between the trigger and associated particles to minimize the near-side jet contributions. 
As seen in Fig.~\ref{fig:corr}, the near-side correlations are mostly consistent with zero within systematic uncertainties. Previous dihadron correlations without $v_3$ subtraction have shown a near-side correlation at large $\deta$ in heavy-ion collisions~\cite{jetspec,ridge}, called the ridge, suggesting the ridge appears to be mainly due to $v_3$. However, there appears a finite ridge remaining for in-plane trigger particles ($\phis<15^\circ$) beyond the maximum flow subtraction. 

Unlike the near side, the away-side correlation is finite for all $\phis$. The correlation structure evolves with trigger particles moving from the in-plane to the out-of-plane direction. The away-side correlation is single peaked, similar to \dAu\ results, for in-plane trigger particles and appears to be significantly broadened or double-peaked for out-of-plane trigger particles. 

The effect of a $\phis$-dependent $v_2$ is investigated~\cite{long} and found to eliminate the ridge correlation entirely. However, the exercise does not reveal the physics mechanism of the possible ridge because the $\phis$-dependent $v_2$ is a manifestation of a $\phis$-dependent ridge, and vice versa. Even with the subtraction of a $\phis$-dependent $v_2$, the away-side structure remains robust~\cite{long}. The possible bias in event-plane reconstruction by the trigger particle and its associated (away-side) particles is investigated and is unlikely to be the cause of the observed away-side structure~\cite{long}.

\begin{figure}
\begin{center}
\includegraphics[width=0.4\textwidth]{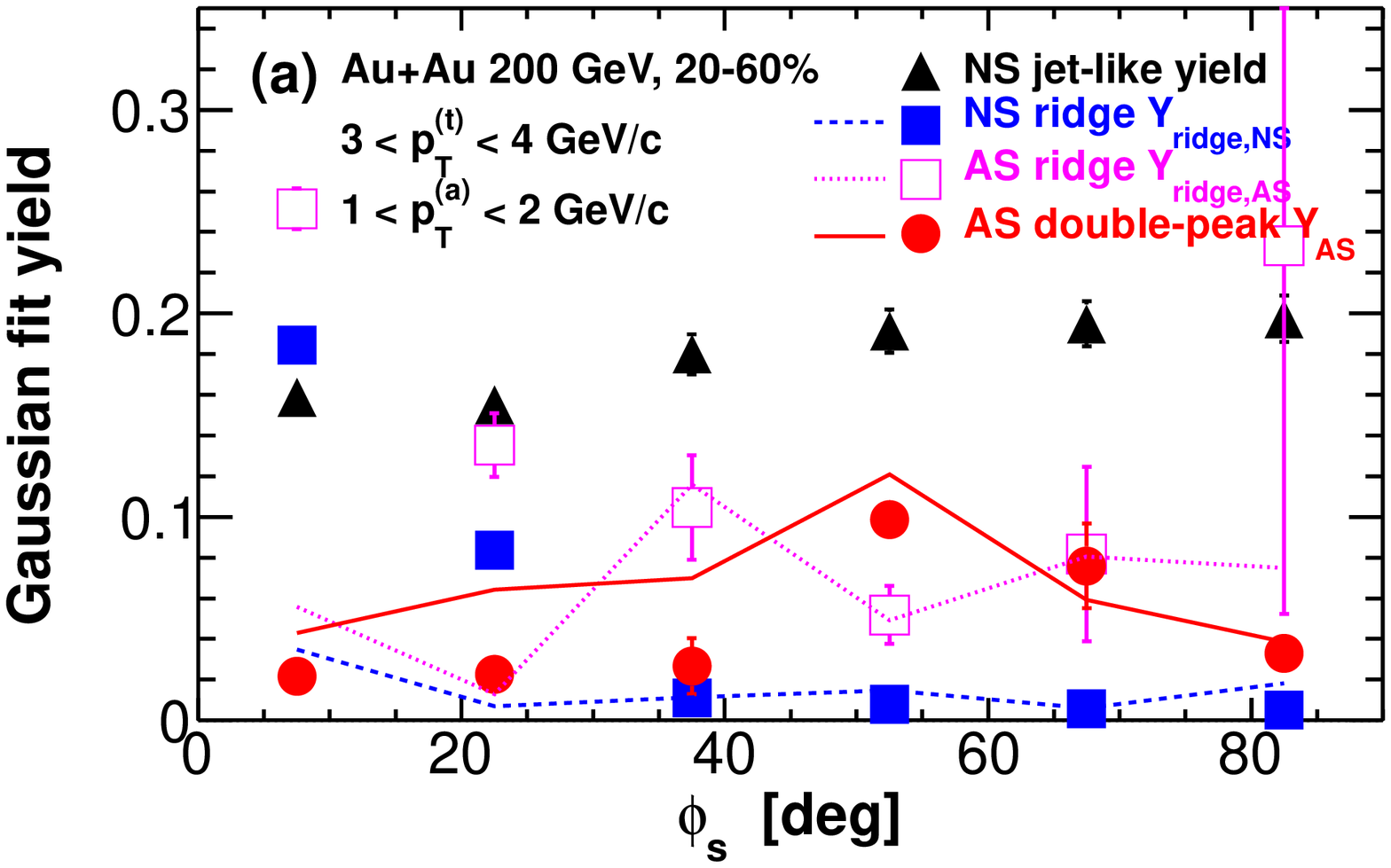}
\includegraphics[width=0.4\textwidth]{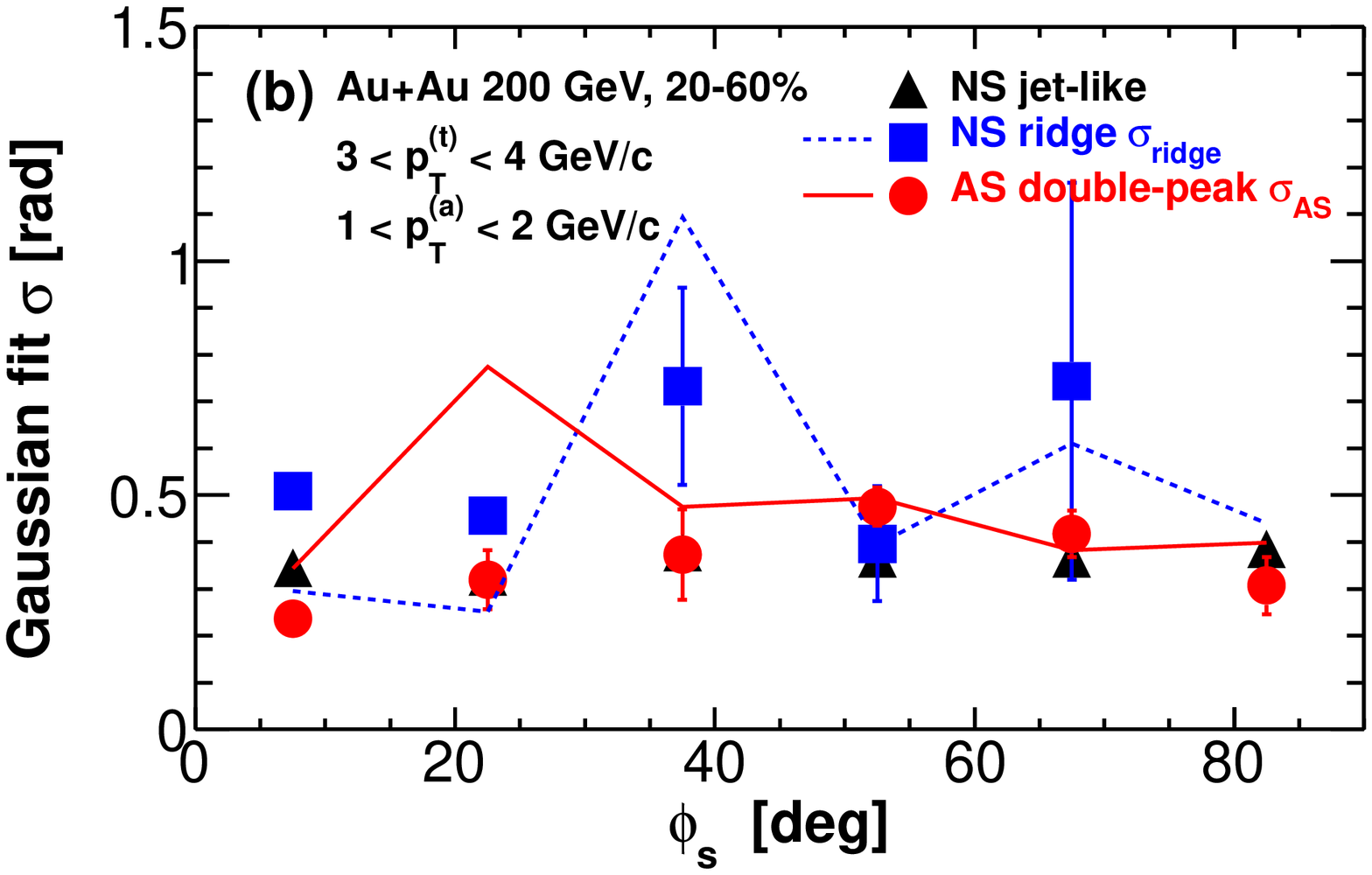}
\includegraphics[width=0.4\textwidth]{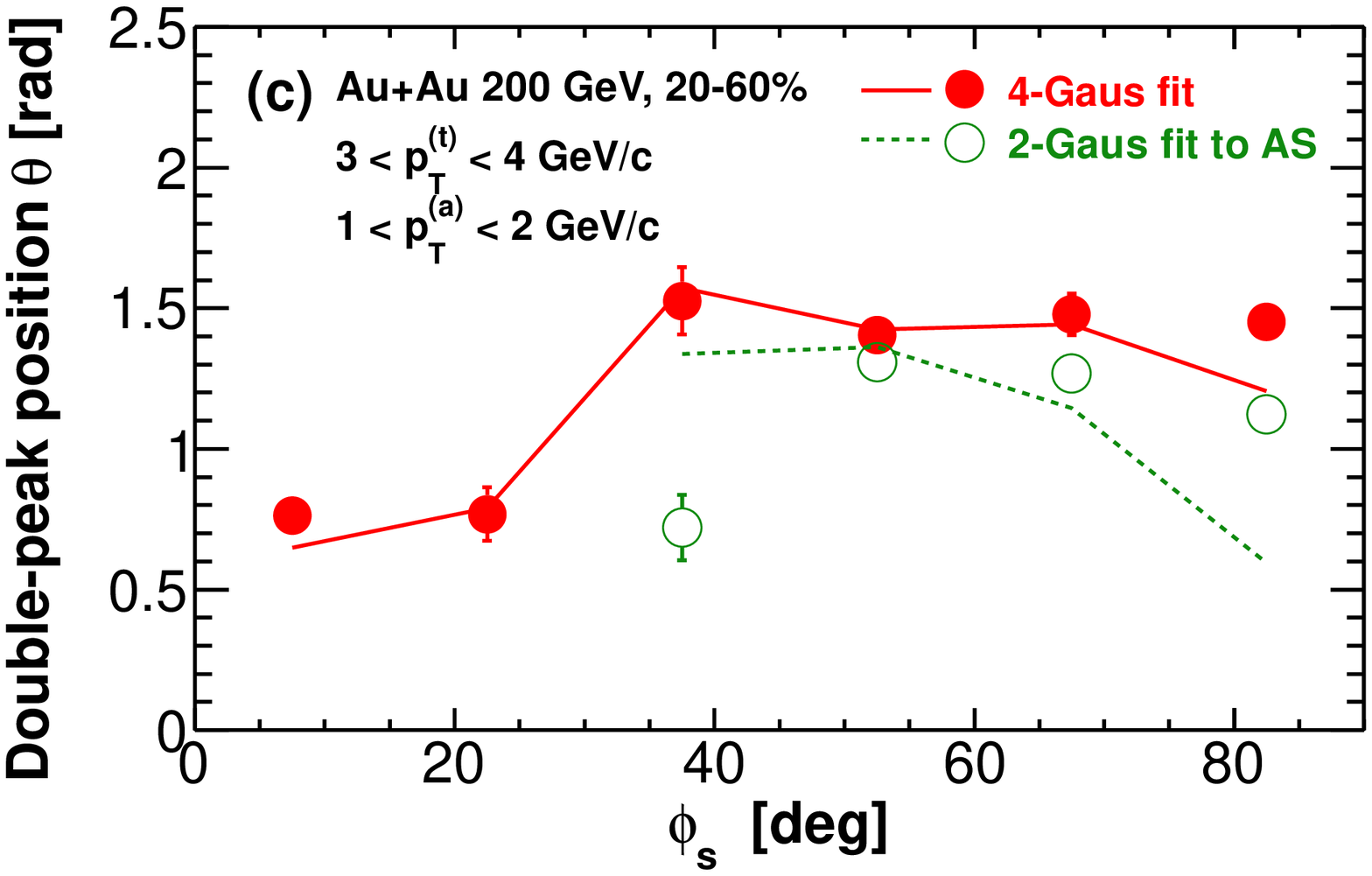}
\end{center}
\caption{(Color online) Parameters of four-Gaussian fit to the background subtracted dihadron correlations at $|\deta|>0.7$ as a function of $\phis$. The near-side (NS) jet-like correlation results are also shown; they are obtained by the difference in $\dphi$ correlations at small and large $\deta$ from Ref.~\cite{long}. (a) Correlated yields where the NS jet-like ($|\deta|<0.7$) and ridge ($|\deta|>0.7$) yields are obtained from bin counting within $|\dphi|<1$; (b) Gaussian peak widths in $\dphi$; and (c) The away-side (AS) double-peak Gaussian centroid. For comparison the centroid from a two-Gaussian fit to the AS correlation ($|\dphi|>0.7)$ is shown for the four out-of-plane slices.
Error bars are statistical only. The curves correspond to the results with maximum flow subtraction by the two-particle cumulant method, which indicate the systematics. 
}
\label{fig:fit}
\end{figure}

To study the structure of the large $\deta$ correlation functions quantitatively, the data are fit with two away-side Gaussian peaks symmetric about $\dphi=\pi$, a near-side Gaussian at $\dphi=0$ for the ridge, and a back-to-back Gaussian at $\dphi=\pi$ (referred to as away-side ridge) with identical width as the near-side ridge~\cite{long}. Namely
\begin{widetext}
\begin{equation}
\frac{1}{N_{\rm trig}}\frac{dN}{d\dphi}=\frac{Y_{\rm AS}}{\sqrt{2\pi}\sigma_{\rm AS}}\left(e^{-\frac{(\dphi-\pi+\theta)^2}{2\sigma_{\rm AS}^2}}+e^{-\frac{(\dphi-\pi-\theta)^2}{2\sigma_{\rm AS}^2}}\right)+\frac{1}{\sqrt{2\pi}\sigma_{\rm ridge}}\left(Y_{\rm ridge,NS}e^{-\frac{(\dphi)^2}{2\sigma_{\rm ridge}^2}}+Y_{\rm ridge,AS}e^{-\frac{(\dphi-\pi)^2}{2\sigma_{\rm ridge}^2}}\right)\,,
\end{equation}
\end{widetext}
where the Gaussians are repeated with period of $2\pi$. The magnitudes of the ridge Gaussians are allowed to vary independently according to the data. The fit parameters are shown in Fig.~\ref{fig:fit} as a function of $\phis$. The data points are results with default $v_2$ subtraction and the curves are the corresponding results with the maximum flow subtraction by the two-particle cumulant $\ff{2}{2}$. Both have subtracted the $v_3$ background using the two-particle cumulant $\ff{3}{2}$. The curves, thus, indicate the results with the maximum systematic uncertainty on one side.

Figure~\ref{fig:fit}(a) shows the Gaussian peak areas of the different correlation components. As a comparison, also shown is the jet-like yield at small $\dphi$ and $\deta$ obtained by the difference between $\dphi$ correlations at small and large $\deta$~\cite{long}. The near-side jet-like ($|\deta|<0.7$) and ridge ($|\deta|>0.7$) yields are obtained from bin counting within $|\dphi|<1$. The bin counting and the fit results are consistent. 
Because the jet-like $\deta$ correlation width is approximately 0.35 (also see Fig.~\ref{fig:fit}(b)), contributions from the tails of the jet-like correlation beyond 0.7 in $\deta$ are negligible.
As seen from Fig.~\ref{fig:fit}(a), the near-side ridge is mostly consistent with zero except in the in-plane direction where a finite ridge beyond the maximum flow systematics seems to be present. The away-side ridge is larger than the near-side ridge at all $\phis$. The double-peak strength appears to increase with $\phis$; for in-plane triggers, where the away-side is single-peaked, there exists a double-peak component if the $\dphi\sim\pi$ region is populated by a Gaussian of the same width as the near-side ridge. 

Fig.~\ref{fig:fit}(b) shows the Gaussian fit widths. The widths do not seem to depend on $\phis$, however, the present systematic uncertainties are large. 
Fig.~\ref{fig:fit}(c) shows the fitted double-peak Gaussian centroid in filled circles.
For the four large $\phis$ bins where the away-side double-peak is observable, the peak location is far removed from $\pi$, almost at $\pi/2$ and $3\pi/2$.
The away-side correlation can also be well fit by only two Gaussians symmetric about $\dphi=\pi$ (without the back-to-back ridge). The fitted double peak positions for the four out-of-plane slices are shown in open circles.
The double-peak correlation structure has been observed before where $v_3$ contributions were not subtracted~\cite{jetspec,PHENIX}. Whether it is an effect of medium excitation by jet-medium interactions over the long away-side pathlength, such as Mach-cone formation~\cite{machcone}, remains an open question. 
There also can be deflection of away-side correlated particles by the collective flow of the medium, especially in the direction perpendicular to the reaction plane~\cite{Betz}. Deflection of correlated particles may have already been seen in three-particle correlations~\cite{3part} where the diagonal peak is stronger than the off-diagonal peak whereas the unsubtracted $v_3$ (and possible Mach-cone emission) should yield the same strength for those peaks. However, in jet-hadron correlations where the trigger jet has significantly larger energy than the trigger particle in this analysis, no deflection of associated particles is observed~\cite{jethadron}.


In summary, dihadron azimuthal correlations at pseudorapidity difference $|\deta|>0.7$ are reported by the STAR experiment for trigger and associated particle $\pt$ ranges of $3<\ptt<4$~\gev\ and $1<\pta<2$~\gev\ in non-central 20-60\% Au+Au collisions as a function of the trigger particle azimuthal angle relative to the event plane, $\phis=|\phit-\psiEP|$. Anisotropic $v_2$, $v_3$, and $v_4$ flow backgrounds are subtracted using the Zero Yield At Minimum (\zyam) method, where the maximum flow parameters are obtained from two-particle cumulant measurements with $\eta$-gap of 0.7. Minimum-bias \dAu\ collision data are presented for comparison. The background subtracted dihadron correlations are found to be modified in Au+Au collisions relative to \dAu; the modification depends on $\phis$. The near-side ridge previously observed in heavy-ion collisions may be largely due to triangular flow $v_3$; After $v_3$ subtraction, however, a finite residual ridge may still be present for in-plane trigger particles. The away-side dihadron correlation broadens from in-plane to out-of-plane, and appears to be double-peaked for out-of-plane trigger particles. The trends of the away-side modification may underscore the importance of pathlength-dependent jet-medium interactions, and should help further the current understanding of high-density QCD in relativistic heavy-ion collisions.

We thank the RHIC Operations Group and RCF at BNL, the NERSC Center at LBNL and the Open Science Grid consortium for providing resources and support. This work was supported in part by the Offices of NP and HEP within the U.S.~DOE Office of Science, the U.S.~NSF, the Sloan Foundation, the DFG cluster of excellence `Origin and Structure of the Universe' of Germany, CNRS/IN2P3, STFC and EPSRC of the United Kingdom, FAPESP CNPq of Brazil, Ministry of Ed.~and Sci.~of the Russian Federation, NNSFC, CAS, MoST, and MoE of China, GA and MSMT of the Czech Republic, FOM and NWO of the Netherlands, DAE, DST, and CSIR of India, Polish Ministry of Sci.~and Higher Ed., Korea Research Foundation, Ministry of Sci., Ed.~and Sports of the Rep.~Of Croatia, Russian Ministry of Sci.~and Tech, and RosAtom of Russia.


\end{document}